\documentclass[10pt]{iopart}
\input epsf.tex
\newcommand{\be}{\begin{equation}}
\newcommand{\ee}{\end{equation}}
\newcommand{\bea}{\begin{eqnarray}}
\newcommand{\eea}{\end{eqnarray}}

\newcommand{\s}{\sigma}

\newcommand{\la}{\langle}
\newcommand{\ra}{\rangle}

\begin{document}

\title{Two leg quantum Ising ladder: A bosonization study of the ANNNI model}

\author{D. Allen\dag, P. Azaria\dag, and   P. Lecheminant\ddag}

\address{  
\dag\ Laboratoire de Physique Th\'eorique des Liquides,
Universit\'e Pierre et Marie Curie, 4 Place Jussieu, 75252 Paris,
France}
\address{
\ddag\ Laboratoire de Physique Th\'eorique et Mod\'elisation,
Universit\'e de Cergy-Pontoise, 5 Mail Gay-Lussac,
Neuville sur Oise, 95301
Cergy-Pontoise Cedex, France}


\begin{abstract}

The quantum ANNNI chain in a transverse field is investigated
by means of the bosonization approach in the limit of large 
next-nearest neighbor interaction. In this regime,
this model can be viewed as a weakly coupled two-leg zigzag ladder which
enables us to derive its low energy effective field theory.
In particular, it is shown that the effect of frustration in the 
system is captured by the presence of a non zero conformal
spin perturbation that accounts for the 
existence of all the incommensurate phases of the model.
\end{abstract}


\maketitle

\nosections
One of the most striking effect of frustration in magnetic systems 
is that it can lead to a huge number of degenerate ground states. 
As a result, the nature
of finite temperature phases may not, in contrast with ferromagnetic
systems, be linked only to the sole nature of the
microscopic degrees of freedom and the dimension of space-time. 
Since frustration
induces strong fluctuations that involve large number of spins,  the low energy
physics is determined by interferences 
between very different regions of phase space.
With this picture in mind one may not expect the field theoretical description
of the frustration
to be an easy  task. Of course there have been several
attempts to describe frustrated magnets by field theories. 
But, to our knowledge
they were mostly restricted to models 
that displayed well defined ground states, such
as helical ordering, and the main 
effect of frustration was captured by an enlargement
of the dimension of the order parameter space\cite{diep}.
Here our aim is
to single out an operator, in the Renormalization Group (RG) sense, 
that captures
the effect of frustration. In this respect, we shall consider
the most studied frustrated system i.e. the two dimensional ANNNI model
(see Refs.\cite{Selke3,yeomansrev,Chakrabarti} for
a review).  This model is characterized by a nearest-neighbor
ferromagnetic interaction ($-J_1 < 0$) and a competiting next-nearest neighbor
antiferromagnetic interaction ($J_2>0$):
\be
{\cal H} = -J_1  \sum_{(i,j)}  \sigma_i \sigma_j + J_2  \sum_{[k,l]}  \sigma_k \sigma_l,
\label{Hannni}
\ee
where $(i,j)$ denotes nearest neighbor 
pairs and $[k,l]$ next-nearest ones along a single space direction.
The phase
diagram of the model (\ref{Hannni}) is very rich and well known and can be
summarize as follows. At zero
temperature, the ground state is ferromagnetic for $J_2 < J_1/2$
and antiferromagnetic 
(actually an antiphase $\uparrow \uparrow \downarrow \downarrow $)  
for  $J_2 > J_1/2$. At the special point
$J_2= J_1/2$, the ground state is infinitely 
degenerate and correlation
functions are short ranged. At finite temperatures ($T\neq 0$), one observes
five different phases depending 
on the parameters (see figure (1)):  ferromagnetic (F), paramagnetic
commensurate (PC), paramagnetic incommensurate (PI), incommensurate critical
phase (IC, also usually called ``floating phase'') 
and antiphase (A). Three different kind of phase transitions are
present in this diagram: An Ising transition between the  F
and PC phases, a commensurate-incommensurate transition\cite{japarid79}
between the A and IC phases, and a Berezinski-Kosterlitz-Thouless
(BKT) transition separates the IC and  PI phases. 
There is also a disorder line that extends down to zero temperature
which divides the PC and PI phases\cite{Peschel}.
As a most notable feature of this
phase diagram is the existence of an incommensurate critical phase 
in a finite
region of the parameter space. 
In a seminal work,
Villain and Bak \cite{villainbak} have proposed an 
approximate effective theory in terms of fermions 
valid in the vicinity of the degenerate point $J_2= J_1/2$. 
All their
predictions have been further confirmed by numerical investigations and series
expansions. 
It is the purpose of the present letter to propose a complementary
and alternative low energy description of the ANNNI model valid
in the large $J_2$ limit. Furthermore, within our approach,
we exhibit 
a particular operator that is at the origin of 
the incommensurate phases found in the phase diagram (figure (1)). 
\begin{figure}[ht]
\begin{center}
\noindent
\epsfxsize=0.6\textwidth
\epsfbox{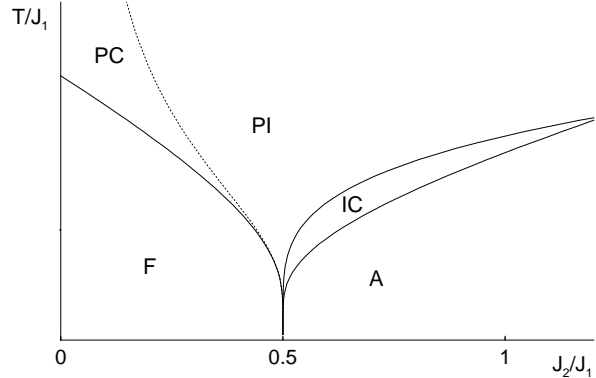}
\end{center}
\caption{\label{phasani}%
Phase diagram of the classical two-dimensional ANNNI model.
}
\end{figure}

\begin{figure}[ht]
\begin{center}
\noindent
\epsfxsize=0.5\textwidth
\epsfbox{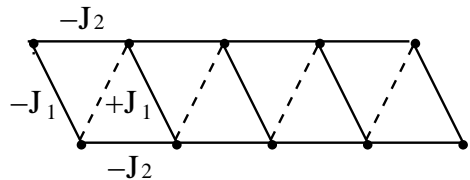}
\end{center}
\caption{\label{zzchain}%
In the large $J_2$ limit, the ANNNI model is better seen
as two weakly coupled quantum Ising chains.
}
\end{figure}

 Our starting point is to map (\ref{Hannni}) into a quantum
Ising model in a transverse magnetic field. This is done by introducing
an anisotropy between the nearest neighbor interaction in the $x$-direction
($J_1$) and in the $y$-direction ($J_\tau$). 
As it is well known (see Ref. \cite{Chakrabarti} for a review), 
the physics
of the classical model (\ref{Hannni}) should be equivalent to the one described
by the following quantum Hamiltonian:
\be
{\cal H}  =  \sum_{n}\left[-\beta J^*_\tau \sigma^z_n
+ \beta J_2 \; \sigma^x_n \sigma^x_{n+2}
-  \beta J_1  \; \sigma^x_n \sigma^x_{n+1}\right],
\label{isingtrans}
\ee
where $\sigma^x_n$, $\sigma^z_n$ are Pauli matrices and 
$2 \beta J^*_\tau = \ln \coth(\beta J_\tau)$.
The model (\ref{isingtrans}) is nothing but the 
one-dimensional $axial$ next-nearest neighbor Ising (ANNNI) model
in a transverse field. As seen in 
figure (\ref{zzchain}), it 
can also be viewed as two quantum Ising chains labelled
(1) and (2) coupled by a ``zigzag'' interaction with 
strength $J_1$ and thus identifies
with a {\sl two-leg quantum Ising ladder} with Hamiltonian:
\bea
{\cal H} &=& \sum_{n,a=1,2}\left[ -\beta J^*_\tau\sigma^z_a
(n+\frac{a}{2}) -\beta J_2\sigma^x_a (n+\frac{a}{2})
\sigma^x_a (n+1+\frac{a}{2}) \right]\nonumber\\
&&-{\beta J_1\over 2}\sum_{n}\sigma^x_1(n+\frac{1}{2})
\left[\sigma^x_2(n)-\sigma^x_2(n+1)\right] +(1\rightarrow 2).
\label{isingladder} 
\eea                                                                   
Notice that in order to obtain (\ref{isingladder}) we have 
performed an (unphysical) gauge transformation on the a{\it th} chain ($a=1,2$):
$\s_a^x(n+a/2)\rightarrow(-1)^{n+a}\s_a^x(n+a/2)$.
As we shall now see,  the model (\ref{isingladder}) can be 
consistently described by a continuous field theory
in the limit $J_1<<J_2$ and $J_2 \sim J^*_\tau$.


{\sl Continuum limit}. 
We shall study the model in the vicinity of the antiferromagnetic phase where
$J_1 << J_2$. One can  take advantage that in the limit
$J_1 = 0$ the model is equivalent to two decoupled Ising models which
are critical when $J_2=J^*_\tau$. One can therefore expand the theory around
the conformal invariant fixed point with symmetry Z$_2 \otimes$ Z$_2$. 
In the critical regime
the two Ising chains are described by two  pairs of 
right and left Majorana (real) fermions
$\psi_{a(R,L)}, a=1,2$:
\be
{\cal H}_0 = -i\frac{v}{ 2} \sum_{a=1,2} \; \left( \psi_{aR}\partial_x\psi_{aR} -
 \psi_{aL}\partial_x\psi_{aL} \right)
-m \; (\epsilon_1 +\epsilon_2),
\ee
$\epsilon_a = i\psi_{aR}\psi_{aL}$ being the energy operator
of the a{\it th} Ising model whereas the mass gap and the velocity are given by:
$m= 2(J_\tau^* - J_2)/T << 1$,  
$v=a_0 J_\tau^*/T$ ($a_0$ being the lattice spacing).
In the absence of interchain interaction, the system is 
disordered (respectively ordered) 
when  $m>0$  (respectively $m<0$) and one has 
$\la\sigma_a\ra = 0$ 
(respectively $\la\sigma_a\ra \neq 0$) 
where $\sigma_a$ 
are the order operators
associated with the two Ising models.
We consider now the interacting case and 
take the continum limit of the second term in equation (\ref{isingladder}) 
in the regime $J_1 \ll J_2$ to obtain:
\be
{\cal H} \simeq {\cal H}_0 + g \;  {\cal V}, \; \; 
{\cal V} = a_0^{1/4}\left( \s_1\partial_x\s_2 -\s_2\partial_x\s_1 \right).
\label{twist}
\ee
When $m=0$ and $g=J_1/2T = 0$, the model is 
conformal invariant with central charge  
$c=\frac{1}{2} +\frac{1}{2} =1$. In the generic case, this fixed point
is perturbed by the thermal operators $\epsilon_{1(2)}$ which have scaling
dimension $\Delta=1$ and the operator ${\cal V}$ which has scaling dimension
$\Delta= 5/4$. It is important to stress that both operators reflect 
very different
physical behaviors. While the former is the standard operator measuring
deviation from criticality in non-frustrated Ising magnets, 
the latter encodes the
whole effect of frustration and is responsible of the non trivial phases
of the ANNNI model as we shall see. 
The operator ${\cal V}$ manifests itself  by the fact that it 
is a parity symmetry breaking perturbation with a conformal spin equals to 1. 
The effect of such non-zero conformal spin term is non trivial
since the usual irrelevant versus relevant criterion does not hold
for such a non-scalar contribution (see for instance the discussion
in Ref. \cite{GNT}). 
In fact, a similar  operator also  appears in the study of the
S=1/2 Heisenberg zigzag ladder\cite{nerses98} and has 
been called a ``twist term''.
Such contribution is difficult to handle non-perturbatively but it has been
stressed that this term may represent a new mechanism 
for incommensuration in one-dimensional systems\cite{nerses98,tsvelik}. 
As we shall see, in the particular case of the ANNNI model, 
the effect of this operator can be elucidated by means of the 
bosonization approach and accounts for the formation of the 
non trivial incommensurate phases depicted in figure (1).

{\sl Bosonization and Effective Theory}. We shall use 
the well known equivalence between two
critical Ising models, characterized by four chiral real fermions
$\psi_{1R(L)}$ and $\psi_{2R(L)}$, and a free boson theory 
described by the chiral fields $\phi_{R(L)}$. 
The bosonization rules are (see Ref. \cite{GNT} for a review):
\be
(\psi_{1}+i\psi_{2})_{R(L)} = \frac{1}{\sqrt{\pi a_0}}
\exp (\pm i\sqrt{4\pi}\phi_{R(L)}),
\ee
from which it follows that ${\cal H}_0$ is equivalent to a sine-Gordon
model at $\beta^2 = 4 \pi$ (free fermion point).
The bosonization of the twist term (\ref{twist}) requires more work. A suitable
bosonic expression is obtained by considering the 
following operator product expansion (OPE) that stems from the fact that 
the order operator $\sigma_a$ is a primary field:
\be
\left(T_1 - T_2\right) (z) \s_1\s_2(w,{\bar w}) \sim {1\over z-w}
\left[ \s_2\partial\s_1 - \s_1\partial\s_2\right](w,{\bar w}),
\label{opedef}
\ee
where $T_a (a=1,2)$ are the energy momentum tensors in 
the holomorphic sector ($z=v\tau+ix$) associated with the Ising
models.
The fields in the left hand side of equation (\ref{opedef}) can be expressed in
terms of the bosonic fields:
$T_1-T_2 = \cos(\sqrt{16\pi}\phi_L)$ and
$\s_1\s_2 = \sqrt 2 \sin(\sqrt\pi \Phi)$ ($\Phi = \phi_L+\phi_R$). 
We thus deduce by
performing the OPE in the bosonic theory the following 
representation of the twist term: 
\be
\fl \quad {\cal H} = \frac{v}{2} \; \left( \left(\partial_x \Phi\right)^2 + 
\left(\partial_x \Theta\right)^2 \right) - \frac{m}{\pi a_0} 
\cos\left(\sqrt{4\pi}\Phi\right)
- \rmi \frac{g\sqrt{2}}{a_0} 
 \cos\left(\sqrt{\pi}\Phi\right)\sin\left(\sqrt{4\pi}\Theta\right),
\label{hboso}
\ee
$\Theta = \phi_L-\phi_R$ being the dual field. 
The next step of our approach is 
to map (\ref{hboso}) onto the XXZ Heisenberg model in a
magnetic field: 
\be
\fl \quad {\cal H} = \frac{ 2\pi u}{3} \;
 \left( {\bf J}_R^2 + {\bf J}_L^2 \right) + 4\pi  m a_0
 \left( J_R^x J_L^x  + J_R^y J_L^y \right) +
\pi g \sqrt{2} \left(  J_R^y -J_L^y \right)
+ 2\pi\lambda_0 J_R^z J_L^z,
\label{hbosocur}
\ee
where $u=5 v/4$, $\lambda_0= 3v/2$ and  the SU(2)$_1$ Kac-Moody
currents are given by $J_{L,R}^z = \partial_x\phi_{L,R}/\sqrt{2 \pi}$ and 
$J^{+}_{L,R} = \exp{(\pm i\sqrt{8\pi}\phi_{L,R})}/(2\pi a_0)$. 
The two expressions  (\ref{hboso}) and  (\ref{hbosocur})
can be shown to be equivalent by a canonical
transformation at the special value $\lambda_0= 3v/2$.
As a result, 
in an appropriate basis, the twist operator acts as 
a magnetic field. 
The  model (\ref{hbosocur}) is more conveniently analyzed in 
a basis where  the magnetic field 
lies along the $z$ axis. To do so, we perform a $\pi/2$ rotation 
around the x-axis in the spin space so that one finally obtains 
after rebosonizing once again the currents:
\be
\fl \; {\cal H} = \frac{u^*}{2} \; \left( \left(\partial_x \Phi\right)^2 + 
\left(\partial_x \Theta\right)^2 \right)
- g_1 \cos\left(\sqrt{8\pi Q}\Phi\right)
 - g_2 \cos\left(\sqrt{8\pi/Q}\Theta\right)
- h  \; \partial_x \Theta,
\label{main}
\ee
where
\be
Q = \sqrt{ \frac{1 - m a_0/u}{1 + m a_0/u}}, \; \; 
u^*  = u\sqrt{(1 + m a_0/u)(1 - ma_0/u)},
\label{Qu*}
\ee
$g_{1(2)} = (2m \pm \lambda_0/a_0)/(4\pi a_0)$,
and $h= g \sqrt{\pi/Q}$. The effective field theory (\ref{main})
is the main result of this work and all the different
phases (apart from the F and PC phases) 
observed in the ANNNI model can be deduced from a 
simple analysis of it.

{\sl Phase Diagram}. Consider first the 
high-temperature phase, i.e. when $m > 0$.
Since $Q < 1$, for sufficiently large $m$, 
the term  $\cos(\sqrt{8\pi/Q}\Theta)$ 
is strongly  irrelevant and can be dropped. The remaining theory is then 
easy to analyse and  a mass gap  to all excitations is generated due to
the presence of the relevant $\cos(\sqrt{8\pi Q}\Phi)$ term. 
On the other hand the operator
$\partial_x\Theta$ leads to incommensurate 
fluctuations of the $\Theta$ dependent correlation functions.
This phase corresponds to the incommensurate paramagnetic phase PI. 
This picture is confirmed
by an exact solution at the special value 
$Q=1/2$ where the model becomes equivalent
to that of free fermions with dispersion 
$\epsilon_{\pm}^2 (k) = 
v^2 \left(k\pm{g \sqrt{2}\pi/v}\right)^2 + 9v^2/16a_0^2$. 
As readily seen,
there is a spectral gap and incommensuration develops as soon as $g\neq 0$
with wavevector $k^*= \sqrt{2}\pi g/v$.

Similarily one can study the low temperature regime i.e. $m<0$ and  $Q>1$. For
sufficiently large $m$, it is now the  $\cos(\sqrt{8\pi Q}\Phi)$ term that 
is irrelevant and can be dropped. After a duality transformation, the resulting 
Hamiltonian is equivalent   to the  XXZ chain  in a magnetic field along the $z$ axis 
which is equivalent to  the bosonized version
of the Villain-Bak theory\cite{Haldane} derived in the vicinity of the degenerate point $ J_{2}/J_{1}
=1/2$, $T=0$.
At small $g$, there is a gap to all excitations with no
incommensuration: it corresponds to the A phase. As $g$ grows,
the magnetic field increases until it reaches the gap 
at some critical value ($g_{c1}$) above which the excitations become massless.
For $g > g_{c1}$, the system displays as well an incommensurate behavior
with wavevector $k^* \sim \sqrt{g^2 - g_{c1}^2}$ and one enters 
the floating phase IC. The nature of the transition at $g=g_{c1}$ is 
of a commensurate-incommensurate type\cite{japarid79}. 
Finally, as $g$ further increases, 
the $\cos({\sqrt{8\pi Q}\Phi})$
term eventually becomes relevant and opens a gap at a critical value $g_{c2}$
where a BKT transition to the PI phase takes place. This picture is, as above,
confirmed by an exact solution at the point $Q=2$ where the model (\ref{main})
becomes equivalent to massive free fermions.

As seen, the previous analysis 
correctly reproduces the phase diagram of the ANNNI
model in the vicinity of the A phase where frustration plays
its tricks. To do so, we have assumed that $|m|$ was sufficiently large
to be able to neglect one of the cosine terms in equation (\ref{main}). 
One can question the validity of this scheme when
$m\sim 0$ where both $\cos(\sqrt{8\pi Q}\Phi)$ and $\cos(\sqrt{8\pi/Q}\Theta)$
operators, being almost marginal, compete. 
A detailed analysis of the RG equations associated
with (\ref{main}) is thus called for 
but one is faced with the difficulty that the coupling constants
are not small since $\lambda_0$ is of order one. One has 
therefore to make the hypothesis that 
the qualitative feature of the RG approach does not depend 
on the strength of $g_1$ and $g_2$ and treat them as small 
couplings. The RG equations  have already 
been obtained  in a different context by
Giamarchi and Schulz\cite{Giamarchi}. It follows from \cite{Giamarchi} that 
the results we have obtained for large $|m|$ remain valid for small $m$
confirming our hypothesis. However, there is room for
a spin-flop transition from the A phase directly to the 
PI phase if the $\cos(\sqrt{8\pi Q}\Phi)$
perturbation blows up before the $\partial_x\Theta$ term closes the gap.
The occurence of such a transition  
strongly depends on the microscopic couplings
of the bare Hamiltonian. 
The analysis of the RG equations for  $\lambda_0 = 6u/5$ reveals
that a spin flop transition does not occur for the ANNNI model. 
However, care has to be payed since for this value of $\lambda_0$,
perturbation theory strictly does not apply and this 
leaves open  the question of
the existence of a Lifshitz point in the ANNNI model. Furthermore,  notice
that the presence of a Lifshitz point is a generic feature 
of (\ref{main}). Therefore one may expect that 
other lattice Hamiltonians displaying the same qualitative
phase diagram than the ANNNI model will exhibit such a point. 
In summary, we have derived a low energy description of the ANNNI model
in the large $J_2$ limit  where the system can be viewed
as a weakly coupled two-leg zigzag ladder. This  enables us to
start from a conformal invariant fixed point in the vicinity of which the continuum
limit is well defined and frustration manifests itself through the twist 
operator. This approach accounts for all the incommensurate phases of the ANNNI model
and
in the low temperature limit matches the Villain-Bak description
near the degenerate point.  

D. Allen is a CNRS post-doctoral fellow
and  thanks l'universit\'e Cergy-Pontoise for hospitality while this
work was initiated.

\end{document}